\documentclass[structabstract]{aa}  
\usepackage{graphicx}
\usepackage{color}
\usepackage{epsfig}
\usepackage{longtable,lscape}
\usepackage{subfigure}
\usepackage[varg]{txfonts}
\usepackage{natbib}
\usepackage[utf8]{inputenc}  
\usepackage{rotating}
\bibpunct{(}{)}{;}{a}{}{,}
\def\mpc{h^{-1} {\rm{Mpc}}}
\def\kms {\rm{km~s^{-1}}}
\def\apj {ApJ}
\def\apjl {ApJL}
\def\apjs {ApJS}
\def\aj {AJ}
\def\aap {A\&A}
\def\mnras {MNRAS}
\def\arcsec{''}

\def\Mpc {\rm Mpc}

\begin{document}
 \title{Galaxy properties in clusters. II. Backsplash Galaxies}


   \author{ H. Muriel\inst{1,2} \and  V. Coenda\inst{1,2},}


   \institute{Instituto de Astronom\'ia
  Te\'orica y Experimental IATE, CONICET, Laprida 922, X5000BGR, C\'ordoba, Argentina 
         \and Observatorio Astron\'omico, Universidad Nacional de C\'ordoba, Laprida 854, X5000BGR, C\'ordoba, Argentina \\
         \email{hernan@mail.oac.uncor.edu; vcoenda@mail.oac.uncor.edu }
              }

   \date{Received ;}
\abstract
   {} 
   {We explore the properties of galaxies on the outskirts of clusters and their dependence on recent dynamical 
   history in order to understand the real impact that the cluster core has on the evolution of galaxies.}
   {We analyse the properties of more than 1000 galaxies brighter than $M_{^{0.1}r}$=-19.6 on the outskirts of 90 clusters
($1 < r/r_{vir} < 2$) in the redshift range $0.05<z<0.10$. Using the line of sight velocity of galaxies relative to the 
cluster’s mean, we selected high and low
velocity subsamples. Theoretical predictions indicate that a significant fraction of
the first subsample  should be backsplash galaxies, that is, objects that have already
orbited near the cluster centre. A significant proportion of the sample of high relative velocity (HV) galaxies seems to be
composed of infalling objects.}
   {Our results suggest that, 
at fixed stellar mass, late-type galaxies in the
low-velocity (LV) sample are systematically older, redder and have formed fewer stars during the last 3 Gyrs 
than galaxies in the HV sample. 
This result is consistent with models that assume that the central
regions of clusters are effective in quenching the star formation by means of processes such as ram pressure stripping
or strangulation. At fixed stellar mass, LV galaxies show some evidence of having higher surface brightness and smaller 
size than HV galaxies. These results are consistent with the scenario where galaxies that have orbited the central
regions of clusters are more likely to suffer tidal effects, producing loss of mass as
well as a re-distribution of matter towards more compact configurations.
Finally, we found a higher fraction of ET galaxies in the LV sample,
supporting the idea that the central region of clusters of galaxies may contribute to the transformation of
morphological types towards earlier types.}
   {}


   \maketitle
   
\section{Introduction}
\label{sec:intro}

The relative importance 
of nature vs. nurture (heredity vs. environment) in the formation and evolution 
of galaxies has been increasingly discussed in recent years. There is consensus that both are important and \citet{DeLucia:2012} 
have shown that 
these two elements of galaxy evolution are inevitably intertwined. 
In the downsizing scenario, stars in more massive galaxies tend to have formed earlier and over 
a short period of time (\citealt{Neistein:2006}). This effect in the early stages of galaxy formation
suggests that the initial conditions (nature) are important.

The strong correlation between the present-day galaxy properties and the environment is well known (see
for instance \citealt{Blanton:2005}). 
In the hierarchical clustering scenario, the structures form from the bottom
up, via the merging of small objects into successively more massive systems. In this scenario,
the environment in which galaxies evolve can vary substantially with time. Extreme 
cases are clusters of galaxies, which favour a great number of physical processes that 
could modify the evolutionary history of galaxies. In this environment, gravitational 
phenomena such as the interactions and merger of galaxies are frequent and can produce morphological 
transformations \citep{Moore:1998}. Moreover, the presence of the hot intra-cluster medium favours 
processes like ram pressure (\citealt{GG:1972}, 
\citealt{Abadi:1999}) or strangulation \citep{Larson:1980}, which can affect both the gaseous content 
of galaxies and their star formation history \citep{Fujita:1999}. Galaxies can also suffer the
stripping of gas, stars and dark matter due to interaction with the cluster
potential \citep{Moore:1999}.
These aspects make clusters of galaxies ideal objects for evaluating the processes involved in the  
evolution of galaxies. Recently it has been suggested that group environment 
also plays a key role in present-day galaxy properties (e.g. \citealt{Wilman09}). \citet{Poggianti09} analysed a
combined sample of nearby and high redshift clusters of galaxies and found that the spiral and S0
fractions have evolved more strongly in less massive clusters, suggesting that morphological evolution since 
redshift 1 is more pronounced in low-mass clusters. \citet{Rasmussen:2012} found that the fraction of star-forming 
group members is suppressed relative to the field, mirroring results for massive clusters (see also \citealt{Mahajan:2013}).

The well-known morphological segregation of galaxies into clusters (\citealt{Dressler:1980}) does not imply that the local 
density or distance to the cluster centre remains unchanged. Since the typical time-scale of galaxy orbits in 
a cluster is of the order of a few Gyrs (\citealt{McCarthy:2008}), these galaxies must have passed close to the centre of 
the cluster since their infall. Moreover, it is to be expected that a fraction of the galaxies 
that have been close to the cluster centre, 
would today be on the outskirts. \citet{Gill05} using high-resolution N-body simulations found that 
approximately half of the galaxies that at present have clustocentric distances between 1 and 2 
virial radii, were in the past close to the cluster centre. These galaxies are usually called 
"backsplash". More recently, \citet{Mahajan11} investigated how, at a given projected clustocentric radius, 
the stellar mass and star formation activity of a galaxy depend on its line-of-sight velocity relative to the 
cluster’s mean velocity ($v_{los}$). Based on N-body simulations, they found that the fraction of backsplash objects depends 
on the velocity threshold: one-third (half) at projected radii slightly greater 
than the virial radius and $|v_{los}| < \sigma$ ($|v_{los}| << \sigma$), where $\sigma$ is the velocity dispersion of the 
cluster. If the backsplash and the infalling populations can be identified separately, the effects on the galaxy properties  
caused by the core of clusters can be distinguished from the preprocessing in groups. 
\citet{Mahajan11} analysed a sample of clusters selected from a 
group catalogue and found that star formation in galaxies is almost 
completely quenched in a single passage through the cluster. The role of the backsplash scenario in the evolution 
of galaxies is also tested in \citet{Rines05}, \citet{Pimbblet06} and \citet{Aguerri10}. 

Based on the backsplash 
scenario, in this paper we address the issue of the influence of cluster environment in galaxy evolution.
For this purpose, we use the sample of galaxies in 
clusters selected by \nocite{Coenda:2009} Coenda \& Muriel (2009, hereafter paper I) plus a new subsample of X-ray clusters 
obtained by cross-correlating the Northern ROSAT All-Sky 
Galaxy Cluster Survey (NORAS, \citet{Bohringer:2000}, hereafter B00) with the Sloan Digital Sky Survey (SDSS, 
\citealt{York:2000}). We have analysed a wide range 
of properties of high and low velocity subsamples of cluster members with clustocentric distances $1<r/r_{vir}<2$ 
in the redshift range $0.05<z<0.10$. The results are analysed in the 
framework of models of the formation and evolution of galaxies.

This paper is organised as follows: in Sec. \ref{sec:sample} we describe the samples of clusters and 
galaxies. Galaxy properties and their dependence on environment are analysed in Sec. \ref{sec:results}. 
In Sec. \ref{sec:conclu} we discuss the results and outline our conclusions.
 
\section{The Sample}
\label{sec:sample}

\subsection{The cluster Sample}

Clusters of galaxies used in this paper were selected from samples identified using
different techniques. Two are based on the X-ray emission produced by the hot gas in the intra-cluster 
medium, the ROSAT-SDSS Galaxy Cluster Survey of Popesso et al. (2004, hereafter P04) and NORAS of B00; 
the other one makes use of the well-known segregation of galaxies, according to which 
clusters are mostly populated by red early-type galaxies and the presence of a dominant 
galaxy in the cluster centre: the MaxBCG Catalogue of Koester et al. 
(2007b, hereafter K07). \nocite{Koester:2007}

The ROSAT-SDSS catalogue of P04 \nocite{Popesso:2004} comprises 114 galaxy clusters detected in the 
ROSAT All Sky Survey (RASS) lying in the area surveyed by the SDSS by February 
2003. This X-ray-selected catalogue includes clusters with masses from 
$10^{12.5}\cal{M}$$_{\odot}$ to $10^{15}\cal{M}$$_{\odot}$ in the redshift range 
$0.002\leq z \leq 0.45$ and X-ray luminosity $L_x(0.1 - 2.4 \rm{keV})\leq \,30.14 \times 10^{44} \rm{erg~s^{-1}}$. 
On the other hand, NORAS covers the celestial region of declination 
$\delta \geq 0^o$ and galactic latitude $|b_{II}|\geq 20^o$. The B00 sample comprises 378 X-ray sources 
that show extended X-ray emission in the first processing of the ROSAT All-Sky Survey with redshift $z\leq0.46$ and 
X-ray luminosity $L_x(0.1 - 2.4 \rm{keV})\leq \,32.56 \times 10^{44} \rm{erg~s^{-1}}$.

In paper I, we selected a X-ray sample of 49 cluster of galaxies from P04 in
the redshift range $0.05<z<0.14$, labeled as C-P04-I. Galaxies in these clusters were identified using the
Main Galaxy Sample (MGS; \citealt{Strauss:2002}) of the Fifth Data Release (DR5, \citealt{dr5}) of SDSS, 
which includes spectroscopic redshifts 
down to a \citet{petro76} magnitude $r=17.77$.  In this paper, we expand the X-ray cluster sample using the 
cross-correlation between NORAS and SDSS. We indentify a subsample from B00, that we labeled as C-B00-I, using 
the MGS of the Seventh Data Release (DR7, \citealt{Abazajian:2009}) 
of SDSS. C-B00-I comprises 55 galaxy clusters in the redshift range $0.05<z<0.14$ (clusters are listed in Table \ref{tb:data}). 

The optical MaxBCG catalogue provides sky locations, photometric redshift 
estimates and richness for 13823 clusters. Details of the 
selection algorithm and catalogue properties are published in 
\citet{Koester:2007} and K07, respectively.
The MaxBCG selection relies on the observation that the galaxy population 
of rich clusters is dominated by bright red galaxies tightly clustered 
in colour (the E/S0 ridgeline). Since these galaxies are old, passively 
evolving stellar populations, their $g-r$ colours closely reflect their 
redshifts. 

Galaxies in optically-selected MaxBCG clusters belong to the C-K07-I sample of Paper I. It consists in selected clusters
from the catalogue of K07, using the MGS of DR5. For the C-K07-I sample, we apply a restriction in the richness, 
selecting clusters with $N_{\rm gal}\geq 20$ in order to have cluster masses comparable to those in the C-P04-I sample.
The C-K07-I sample comprises 209 regular galaxy clusters in the redshift range $0.05<z<0.14$.

The selection of cluster members and the estimation of the physical properties of C-B00-I clusters are made
using the same method that we have applied to C-P04-I and C-K07-I in Paper I. To select cluster members, 
we use the friends-of-friends (\textit{fof}) algorithm developed by \citet{H&G:1982} with percolation linking length values 
according to \citet{Diaz:2005}: density contrast of $\delta \rho/\rho=315$ and line-of-sight linking 
length of $V_0=200 \kms$ and a projected separation $D_0=0.189 \Mpc$. As a result, we obtain a list of substructures for each field with at least 10 members
identified by fof. The second step consists of an eyeball examination of the structure detected 
by \textit{fof}. From the redshift distribution of galaxies, we determine
the line-of-sight extension of each cluster and therefore the cluster
members. Through visual inspection, we exclude systems that have two or more close substructures of similar
size in the plane of the sky and/or in the redshift distribution.
Once the members of each cluster are selected, we compute 
cluster physical properties using methods described by \citet{Beers:1990}. These include the line-of-sight velocity 
dispersion $\sigma$, virial radius and mass.

\begin{figure}[t]
   \centering
   \includegraphics[width=9cm]{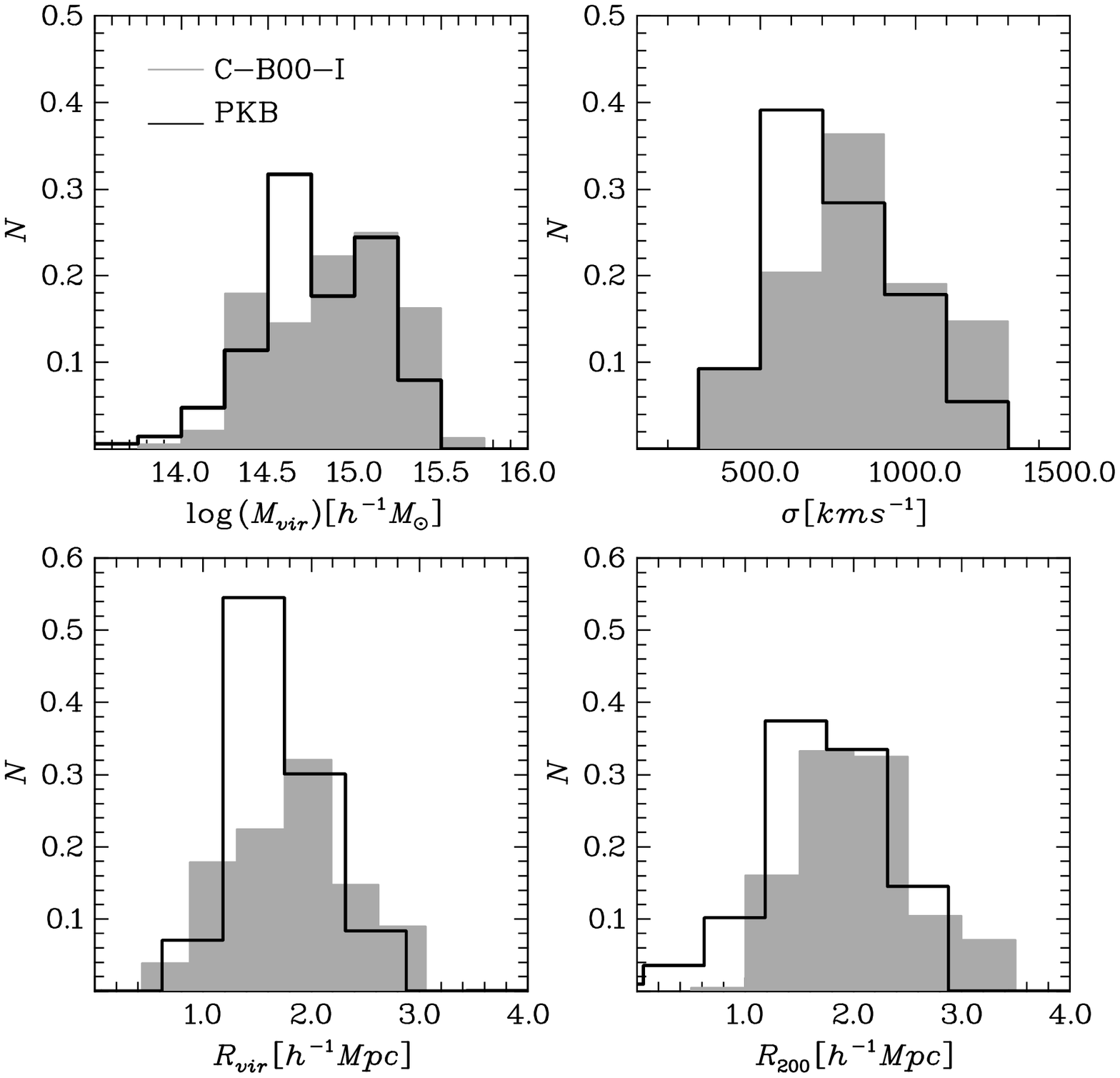}
   \caption{The distributions of cluster physical properties for PKB sample (\emph{black lines}) in the 
redshift range $0.05<z<0.10$ and the C-B00-I sample (\emph{grey lines}) in the redshift range $0.05<z<0.14$.}
\label{fig:fmasa}
\end{figure}

The final sample used in this paper is a composite of 27 C-P04-I, 
32 C-K07-I and 31 C-B00-I cluster of galaxies, that we labeled as PKB, that comprises 90 clusters in the redshift 
range $0.05<z<0.10$. Fig. \ref{fig:fmasa} shows the distributions of cluster physical properties (\emph{black lines})
of our final sample and their median values are shown in Table \ref{tb:mean}.

\begin{table*}
\center
\begin{tabular}{cclllll}
\hline \hline
 $\alpha$ (J2000.0) & $\delta$ (J2000.0) & $z$ & $\sigma$ & $\cal{M}$$_{vir}$ &  $R_{vir}$ &  $R_{200}$ \\
    $[h\ m\ s]$                 &  [$^{o}$ ' '']               &     & [$\kms$] & [$h^{-1}\cal{M}$$_{\odot}$] &[$\mpc$] &[$\mpc$]  \\
\hline
$ 07\,\, 25\,\, 50.8  $ &$ +41\,\, 22\,\, 52.7 $ & $ 0.110 $ & $  430.52 $ & $  0.231\times 10^{15} $ & $ 1.783 $ & $ 1.065 $ \\
$ 07\,\, 46\,\, 37.3  $ &$ +31\,\, 00\,\, 49.0 $ & $ 0.058 $ & $  707.25 $ & $  0.385\times 10^{15} $ & $ 1.103 $ & $ 1.750 $ \\
$ 07\,\, 59\,\, 42.8  $ &$ +54\,\, 00\,\, 05.8 $ & $ 0.102 $ & $  472.30 $ & $  0.224\times 10^{15} $ & $ 1.437 $ & $ 1.169 $ \\
$ 08\,\, 05\,\, 43.3  $ &$ +45\,\, 41\,\, 25.1 $ & $ 0.135 $ & $  974.78 $ & $  0.889\times 10^{15} $ & $ 1.341 $ & $ 2.412 $ \\
$ 08\,\, 11\,\, 02.4  $ &$ +16\,\, 44\,\, 05.6 $ & $ 0.092 $ & $  734.88 $ & $  0.480\times 10^{15} $ & $ 1.275 $ & $ 1.818 $ \\
$ 08\,\, 21\,\, 01.8  $ &$ +07\,\, 51\,\, 57.6 $ & $ 0.110 $ & $  454.99 $ & $  0.210\times 10^{15} $ & $ 1.452 $ & $ 1.126 $ \\
$ 08\,\, 28\,\, 38.9  $ &$ +30\,\, 25\,\, 40.8 $ & $ 0.050 $ & $  827.01 $ & $  0.740\times 10^{15} $ & $ 1.552 $ & $ 2.046 $ \\
$ 08\,\, 44\,\, 37.5  $ &$ +42\,\, 59\,\, 10.0 $ & $ 0.054 $ & $  375.80 $ & $  0.915\times 10^{14} $ & $ 0.928 $ & $ 0.930 $ \\
$ 09\,\, 24\,\, 14.0  $ &$ +14\,\, 09\,\, 52.9 $ & $ 0.140 $ & $  531.81 $ & $  0.357\times 10^{15} $ & $ 1.811 $ & $ 1.316 $ \\
$ 10\,\, 06\,\, 34.2  $ &$ +25\,\, 55\,\, 32.2 $ & $ 0.117 $ & $  469.32 $ & $  0.199\times 10^{15} $ & $ 1.292 $ & $ 1.161 $ \\
$ 10\,\, 16\,\, 43.5  $ &$ +24\,\, 46\,\, 07.7 $ & $ 0.173 $ & $ 1229.69 $ & $  0.287\times 10^{16} $ & $ 2.719 $ & $ 3.043 $ \\
$ 10\,\, 19\,\, 59.8  $ &$ +40\,\, 59\,\, 14.3 $ & $ 0.092 $ & $  901.98 $ & $  0.726\times 10^{15} $ & $ 1.280 $ & $ 2.232 $ \\
$ 10\,\, 31\,\, 35.0  $ &$ +35\,\, 03\,\, 17.6 $ & $ 0.122 $ & $  512.04 $ & $  0.191\times 10^{15} $ & $ 1.046 $ & $ 1.267 $ \\
$ 10\,\, 34\,\, 45.0  $ &$ +30\,\, 41\,\, 17.5 $ & $ 0.137 $ & $  744.20 $ & $  0.641\times 10^{15} $ & $ 1.659 $ & $ 1.841 $ \\
$ 10\,\, 40\,\, 43.9  $ &$ +39\,\, 56\,\, 53.2 $ & $ 0.137 $ & $ 1291.03 $ & $  0.218\times 10^{16} $ & $ 1.877 $ & $ 3.194 $ \\
$ 11\,\, 09\,\, 18.8  $ &$ +41\,\, 33\,\, 44.6 $ & $ 0.076 $ & $  640.30 $ & $  0.353\times 10^{15} $ & $ 1.233 $ & $ 1.584 $ \\
$ 11\,\, 11\,\, 38.5  $ &$ +40\,\, 50\,\, 32.6 $ & $ 0.075 $ & $  742.21 $ & $  0.594\times 10^{15} $ & $ 1.545 $ & $ 1.836 $ \\
$ 11\,\, 22\,\, 14.5  $ &$ +67\,\, 12\,\, 46.4 $ & $ 0.055 $ & $  329.25 $ & $  0.673\times 10^{14} $ & $ 0.890 $ & $ 0.815 $ \\
$ 11\,\, 23\,\, 13.2  $ &$ +19\,\, 35\,\, 58.6 $ & $ 0.104 $ & $  784.16 $ & $  0.810\times 10^{15} $ & $ 1.888 $ & $ 1.940 $ \\
$ 11\,\, 43\,\, 45.5  $ &$ +46\,\, 20\,\, 56.4 $ & $ 0.114 $ & $  496.34 $ & $  0.290\times 10^{15} $ & $ 1.688 $ & $ 1.228 $ \\
$ 11\,\, 55\,\, 18.5  $ &$ +23\,\, 24\,\, 27.0 $ & $ 0.143 $ & $ 1257.93 $ & $  0.331\times 10^{16} $ & $ 3.002 $ & $ 3.113 $ \\
$ 11\,\, 56\,\, 57.8  $ &$ +24\,\, 15\,\, 29.2 $ & $ 0.139 $ & $  899.86 $ & $  0.118\times 10^{16} $ & $ 2.096 $ & $ 2.227 $ \\
$ 12\,\, 01\,\, 47.6  $ &$ +58\,\, 02\,\, 03.8 $ & $ 0.103 $ & $  836.92 $ & $  0.100\times 10^{16} $ & $ 2.049 $ & $ 2.071 $ \\
$ 12\,\, 03\,\, 00.5  $ &$ +28\,\, 34\,\, 50.9 $ & $ 0.133 $ & $  739.39 $ & $  0.312\times 10^{15} $ & $ 0.818 $ & $ 1.830 $ \\
$ 12\,\, 10\,\, 30.6  $ &$ +05\,\, 21\,\, 22.3 $ & $ 0.077 $ & $  642.15 $ & $  0.410\times 10^{15} $ & $ 1.424 $ & $ 1.589 $ \\
$ 12\,\, 25\,\, 12.0  $ &$ +32\,\, 13\,\, 44.8 $ & $ 0.061 $ & $  501.81 $ & $  0.240\times 10^{15} $ & $ 1.365 $ & $ 1.242 $ \\
$ 12\,\, 27\,\, 28.0  $ &$ +08\,\, 49\,\, 44.4 $ & $ 0.090 $ & $  952.88 $ & $  0.123\times 10^{16} $ & $ 1.941 $ & $ 2.358 $ \\
$ 12\,\, 29\,\, 59.2  $ &$ +11\,\, 47\,\, 20.8 $ & $ 0.085 $ & $  878.40 $ & $  0.125\times 10^{16} $ & $ 2.331 $ & $ 2.173 $ \\
$ 13\,\, 03\,\, 45.6  $ &$ +19\,\, 16\,\, 17.4 $ & $ 0.064 $ & $  715.15 $ & $  0.443\times 10^{15} $ & $ 1.240 $ & $ 1.770 $ \\
$ 13\,\, 39\,\, 34.8  $ &$ +18\,\, 30\,\, 43.9 $ & $ 0.114 $ & $  957.25 $ & $  0.130\times 10^{16} $ & $ 2.028 $ & $ 2.369 $ \\
$ 13\,\, 41\,\, 48.8  $ &$ +26\,\, 22\,\, 45.1 $ & $ 0.075 $ & $  684.31 $ & $  0.566\times 10^{15} $ & $ 1.732 $ & $ 1.693 $ \\
$ 13\,\, 48\,\, 53.0  $ &$ +26\,\, 35\,\, 44.2 $ & $ 0.063 $ & $  860.93 $ & $  0.105\times 10^{16} $ & $ 2.038 $ & $ 2.130 $ \\
$ 13\,\, 49\,\, 21.6  $ &$ +28\,\, 06\,\, 13.0 $ & $ 0.076 $ & $  820.02 $ & $  0.108\times 10^{16} $ & $ 2.298 $ & $ 2.029 $ \\
$ 13\,\, 51\,\, 45.6  $ &$ +46\,\, 22\,\, 00.5 $ & $ 0.062 $ & $  571.32 $ & $  0.345\times 10^{15} $ & $ 1.518 $ & $ 1.414 $ \\
$ 13\,\, 54\,\, 08.6  $ &$ +14\,\, 51\,\, 01.1 $ & $ 0.126 $ & $  621.21 $ & $  0.313\times 10^{15} $ & $ 1.161 $ & $ 1.537 $ \\
$ 14\,\, 13\,\, 34.0  $ &$ +43\,\, 40\,\, 24.2 $ & $ 0.089 $ & $  612.68 $ & $  0.310\times 10^{15} $ & $ 1.184 $ & $ 1.516 $ \\
$ 14\,\, 15\,\, 51.6  $ &$ +00\,\, 15\,\, 32.0 $ & $ 0.126 $ & $  466.11 $ & $  0.111\times 10^{15} $ & $ 0.731 $ & $ 1.153 $ \\
$ 14\,\, 21\,\, 35.5  $ &$ +49\,\, 33\,\, 06.8 $ & $ 0.072 $ & $  639.39 $ & $  0.374\times 10^{15} $ & $ 1.311 $ & $ 1.582 $ \\
$ 14\,\, 23\,\, 52.4  $ &$ +40\,\, 15\,\, 42.8 $ & $ 0.082 $ & $  433.13 $ & $  0.156\times 10^{15} $ & $ 1.189 $ & $ 1.072 $ \\
$ 14\,\, 28\,\, 28.1  $ &$ +56\,\, 52\,\, 58.4 $ & $ 0.107 $ & $  791.82 $ & $  0.934\times 10^{15} $ & $ 2.135 $ & $ 1.959 $ \\
$ 14\,\, 31\,\, 07.1  $ &$ +25\,\, 38\,\, 19.7 $ & $ 0.096 $ & $  880.42 $ & $  0.987\times 10^{15} $ & $ 1.826 $ & $ 2.178 $ \\
$ 14\,\, 42\,\, 18.4  $ &$ +22\,\, 18\,\, 17.3 $ & $ 0.096 $ & $  568.21 $ & $  0.402\times 10^{15} $ & $ 1.783 $ & $ 1.406 $ \\
$ 14\,\, 53\,\, 07.2  $ &$ +21\,\, 53\,\, 40.9 $ & $ 0.117 $ & $  920.46 $ & $  0.152\times 10^{16} $ & $ 2.571 $ & $ 2.278 $ \\
$ 14\,\, 54\,\, 31.4  $ &$ +18\,\, 38\,\, 31.2 $ & $ 0.059 $ & $  645.70 $ & $  0.501\times 10^{15} $ & $ 1.724 $ & $ 1.598 $ \\
$ 15\,\, 10\,\, 11.7  $ &$ +33\,\, 30\,\, 52.9 $ & $ 0.113 $ & $ 1214.50 $ & $  0.249\times 10^{16} $ & $ 2.417 $ & $ 3.005 $ \\
$ 15\,\, 11\,\, 06.3  $ &$ +18\,\, 02\,\, 37.3 $ & $ 0.115 $ & $  626.88 $ & $  0.296\times 10^{15} $ & $ 1.081 $ & $ 1.551 $ \\
$ 15\,\, 18\,\, 45.6  $ &$ + 6\,\, 13\,\, 52.3 $ & $ 0.102 $ & $  799.14 $ & $  0.925\times 10^{15} $ & $ 2.076 $ & $ 1.977 $ \\
$ 15\,\, 20\,\, 40.3  $ &$ +48\,\, 39\,\, 37.4 $ & $ 0.073 $ & $  743.10 $ & $  0.543\times 10^{15} $ & $ 1.410 $ & $ 1.839 $ \\
$ 15\,\, 39\,\, 48.7  $ &$ +30\,\, 43\,\, 02.3 $ & $ 0.097 $ & $  778.23 $ & $  0.668\times 10^{15} $ & $ 1.582 $ & $ 1.926 $ \\
$ 15\,\, 58\,\, 20.6  $ &$ +27\,\, 13\,\, 36.8 $ & $ 0.089 $ & $ 1118.92 $ & $  0.248\times 10^{16} $ & $ 2.846 $ & $ 2.769 $ \\
$ 16\,\, 20\,\, 31.7  $ &$ +29\,\, 53\,\, 43.1 $ & $ 0.097 $ & $  866.98 $ & $  0.108\times 10^{16} $ & $ 2.058 $ & $ 2.145 $ \\
$ 16\,\, 54\,\, 23.3  $ &$ +23\,\, 34\,\, 11.6 $ & $ 0.058 $ & $  519.73 $ & $  0.289\times 10^{15} $ & $ 1.533 $ & $ 1.286 $ \\
$ 17\,\, 09\,\, 48.8  $ &$ +34\,\, 26\,\, 26.2 $ & $ 0.086 $ & $ 1094.27 $ & $  0.181\times 10^{16} $ & $ 2.168 $ & $ 2.708 $ \\
$ 17\,\, 17\,\, 57.4  $ &$ +32\,\, 32\,\, 33.4 $ & $ 0.107 $ & $  959.81 $ & $  0.111\times 10^{16} $ & $ 1.729 $ & $ 2.375 $ \\
$ 17\,\, 22\,\, 18.7  $ &$ +30\,\, 44\,\, 08.2 $ & $ 0.047 $ & $  576.22 $ & $  0.271\times 10^{15} $ & $ 1.172 $ & $ 1.426 $ \\
\hline
\end{tabular}
\caption{C-B00-I cluster sample.}
\label{tb:data}
\end{table*}

\begin{table*}[th]
\center
\begin{tabular}{ccccccc}
\hline \hline
  Sample      & $\sigma$     & $R_{200}$ & $\cal{M}$$_{vir}$        & $R_{vir}$     & Redshift    & $\#$ Cluster      \\
              & [$\kms$]     & [$\mpc$]  & [$h^{-1}\cal{M}$$_{\odot}$] &[$\mpc$]                       \\
 \hline
 C-B00-I      &              &            &                                       &                & $0.05 \le z \leq 0.14$ & 55     \\
 Median       &    $820$     &  $2.03$    &      $9\times10^{14}$                 &   $1.83$       &                      &           \\
 Range        & $329-1291$   & $0.81-3.19$&     $7\times10^{13} -3\times10^{15}$  &   $0.73-3.00$  &                      &           \\
 \hline
 C-P04-I      &              &            &                                       &                & $0.05 \le z \leq 0.14$ & 49     \\
 Median      & $715$        & $1.77$     &      $7\times10^{14}$                 & $1.75$          &                        &         \\
 Range        & $332-1060$   & $0.82-2.62$&     $5\times10^{13} -2\times10^{15}$  &   $0.71-2.56$  &                        &      \\ 
 \hline
 C-K07-I      &              &            &                                       &               & $0.05 \le z \leq 0.14$  & 207      \\
 Median      & $675$        & $1.67$     &       $6\times10^{14}$                & $1.59$         &                        &         \\
 Range        & $212-1272$   & $0.53-3.15$&     $3\times10^{13} -2\times10^{15}$  &   $0.76-2.88$ &                        &         \\
 \hline 
 PKB          &              &            &                                       &               & $0.05 \le z \leq 0.10$  & 90      \\
 Median       &  $707$       & $1.75$     &   $6\times10^{14}$                    & $1.64$           &                        &         \\
 Range        & $231-1119$   & $0.53-2.71$&     $3\times10^{13} -3\times10^{15}$  &   $0.71-2.89$    &                        &         \\ 
\hline \hline
\end{tabular}
\caption{Sample's name (column 1), cluster physical properties (columns 2-5), redshift ranges (column 6) and 
total number of clusters (column 7).}
\label{tb:mean}
\end{table*}

\subsection{The galaxy sample}
\label{sec:galsample}

Properties of galaxies can be affected by the environment. Among the most commonly studied are morphology, 
colors, luminosities and sizes. Throughout this work we use two sets of galaxy properties. The first set  
is based on the photometric properties of galaxies: the Petrosian $r$-band absolute magnitude 
$M_{^{0.1}r}$, the radius that 
encloses 50\% of the Petrosian flux $r_{50}$, the $r$-band surface brightness, $\mu_{50}$, computed inside $r_{50}$ and
the $^{0.1}(g-r)$. Colours are computed using model instead of Petrosian magnitudes. Model magnitudes 
are calculated using the best-fit parameters in the $r$-band, and the same model is applied to the other bands; 
the light is therefore measured consistently through the same aperture in all bands.
We exclude galaxies with $r_{50}<2\arcsec$ in order to reduce 
the effect of the PSF on galaxy sizes. Moreover, Petrosian quantities are not affected by seeing effects in 
our redshift range \citep{HydeBernandi:2009}. 
In addition, we correct the magnitudes and sizes by sky level using SDSS simulated galaxies 
(http://www.sdss.org/dr6/products/catalogs/index.html). Finally, we estimate the rest-frame radius $r_{50}$ 
in $r$-band as in \cite{HydeBernandi:2009}. The maximum correction in magnitude and size are 0.03 and 0.1 
respectively. For more details see Paper I. It should be noted that we are using galaxies 
from different SDSS data releases. Based on the information in the SDSS documentation pages and a test with a 
random sample of galaxies we found that the impact of this inhomogeneity in the results of our work are minimum.

As in Paper I, in the analyses below, we classify galaxies into early and late types
according to their $r$-band concentration parameter defined as the ratio
between the radii that encloses 90\% and 50\% of the Petrosian
flux, $C = r_{90}/r_{50}$. The concentration index is known to correlate with the morphological type
and can be used to separate early- (E, S0, and Sa) 
from late-type (Sb, Sc, and Irr) galaxies (\citealt{Shima:2001}, \citealt{Strateva:2001}, \citealt{Park:2005}, 
\citealt{Deng:2013}). Typically, early-type galaxies have $C > 2.5$, while for late-types $C < 2.5$. 

To perform a fair comparison, we weight each galaxy in our
computations by 1/Vmax \citep{Schmidt:68} in order to compensate for the limitations of our flux-limited sample.
Galaxy magnitudes used throughout this paper have been corrected for galactic extinction 
following \citet{sch98}, and absolute magnitudes have been computed assuming  $\Omega_0=0.3$, $\Omega_{\Lambda}=0.7$ 
and $H_0=70~h~{\rm km~s^{-1}~Mpc^{-1}}$ and $K-$corrected using the method of \citet{Blanton:2003}~({\small 
KCORRECT} version 4.1). All magnitudes are in the AB system.

\subsection{Ages, metallicities and stellar masses}
\label{sec:age}

Spectral synthesis codes compare galaxy data with models by combining libraries of stellar spectra  
and prescriptions for the star formation and chemical histories. The implementation of these codes allows 
to estimate the stellar masses, the mean ages 
($\tau$) and metallicities ($Z$) of the stellar population of galaxies. \citet{Rakos:2007} analysed 
galaxies in Coma and Abell 1185, finding a clear correlation between $\tau$ and $Z$ with both 
the stellar mass of galaxies and the environment. They found a significant correlation between galaxy mean 
age and clustocentric distance, such that older galaxies inhabit the core. 
\citet{Ellison:2009} analysed a large sample of cluster galaxies with nebular metallicities and found that galaxies in 
clusters have higher values of $Z$ relative to similar galaxies in low-density environments.
They also found that galaxies in locally rich environments have higher
median metallicities than those in locally poor environments.
Although they do not find any correlation between $Z$ and the cluster properties, their results suggest 
that galaxies have higher metallicities in the core of clusters than on the outskirts.

\citet{Tojeiro09} applied a stellar population model to the SDSS final data 
release using the VESPA code (\citealt{Tojeiro07}). This code is based on the construction of galaxy 
histories from synthetic models. For the modeling of the single stellar population (SSP), the authors use the models of \citet{B&C03} 
and \citet{Maraston05}. For the dust modeling, the code assumes that stars of all ages are affected in the 
same way and two different dust extinction models are considered. For the whole sample of cluster members, we use 
the VESPA code to compute the stellar mass, the mass-weighted age and the mass-weighted metallicity. The
last two quantities are computed using the following equations of \citet{Tojeiro10}: 
\begin{equation}
 \left< { t }_{ age } \right> =\frac { \sum _{ i }^{  }{ { t }_{ i }{ x }_{ i } }  }{ \sum _{ i }^{  }{ { x }_{ i } }  } 
\end{equation}
\begin{equation}
 \left< { Z } \right> =\frac { \sum _{ i }^{  }{ { Z }_{ i }{ x }_{ i } }  }{ \sum _{ i }^{  }{ { x }_{ i } }  } 
\end{equation}
where $t_i$ ($Z_i$) is the mean age (metallicity) and $x_i$ is the star formation fraction in the age bin $i$. 
In order to test if star formation is induced during the infall of galaxies into the clusters, we also compute the star 
formation fraction (SFF) during the last 3 Gyr. For our analysis, we adopt the RunID 2 of \citet{Tojeiro09}, which
considers the SSP of \citet{B&C03} and a two-parameter model for the dust. 

\subsection{The selection of Backsplash Galaxies}

We have stacked all the cluster members in the plane $r/r_{vir}$-$|v_{los}|/\sigma$ where the clustocentric distance $r$ 
is normalised to $r_{vir}$ and the absolute line of sight velocity (relative to the cluster) $|v_{los}|$ is normalised
to the cluster velocity dispersion $\sigma$. \citet{Gill05} suggested that 50\% of the galaxies in the interval 1-2 
virial radii of clusters
are backsplash and typically have $v_{los} < 400 km s^{-1}$. \citet{Mahajan11} worked with normalised velocities
and found that backsplash fractions are approximately one-third at projected radii 
slightly greater than the virial radius and $|v_{los}| < \sigma$. The fraction grows to 0.5 if galaxies with 
$|v_{los}| << \sigma$ are considered. Based on these predictions, we select two samples of galaxies 
in the range $1<r/r_{vir}<2$: galaxies having  $|v_{los}| < 0.5\sigma$ 
(hereafter low-velocity or LV) and galaxies with $|v_{los}| > \sigma$ (hereafter high-velocity or HV). 
Each of these samples has been divided into early (ET) and late-type (LT) using the criterion mentioned in 
section \ref{sec:galsample}. 
An important fraction of the galaxies in the LV sample are expected to have had one or more passes through the 
core of the cluster, while infalling galaxies should be common in the HV sample. The total number of galaxies 
considered in this work is 1060. The number of galaxies on each subsample is quoted in table \ref{tb:gal}.

\begin{table}[h]
\center
\begin{tabular}{ccccc}
\hline \hline
   ET-LV & LT-LV & ET-HV & LT-HV  & TOTAL\\        
\hline
   454   & 187   & 275   & 144   & 1060\\
\hline \hline
\end{tabular}
\caption{Number of galaxies in each of the subsamples analysed and the total number of galaxies.}
\label{tb:gal}
\end{table}

\section{Results}
\label{sec:results}

\begin{figure}[t]
    \centering
   \includegraphics[width=8cm]{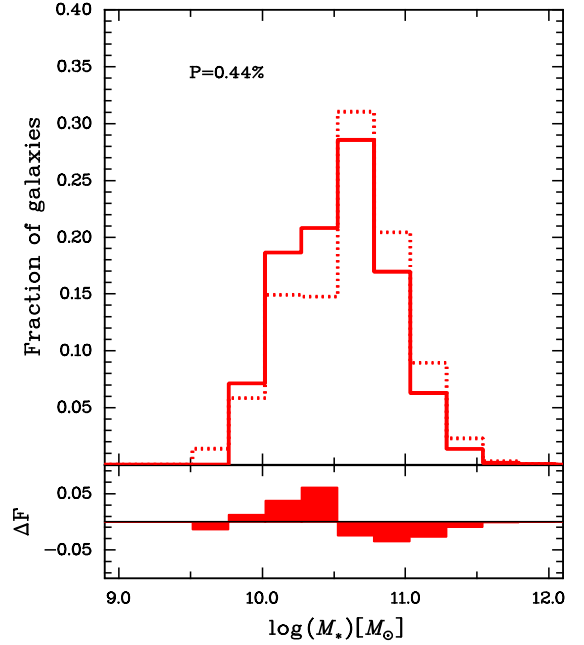}
    \caption{Distributions of the stellar mass for early-type galaxies.  
Continuous line shows the LV galaxies, while HV galaxies are 
shown as dotted line. The \emph{bottom} panel 
shows as {\em shaded histograms} the residuals between the distributions.}
\label{fig:distri_b}
\end{figure}

In Fig. \ref{fig:distri_b} we compare the normalised distributions of the stellar 
mass $M_{\ast}$ for early type galaxies. The continuous line
shows the LV galaxies while the dotted line shows the HV galaxies. In the bottom panel of 
Fig. \ref{fig:distri_b}, we show as {\em shaded histogram} the residuals between the pair of distributions: the difference 
$\Delta F(X)=f_{LV}(M_{\ast})-f_{HV}(M_{\ast})$, where $f_{LV}(M_{\ast})$
and $f_{HV}(M_{\ast})$ are the fraction of galaxies in the bin centred on $M_{\ast}$ in the LV and the HV sample, respectively.
Based on a Kolmogorov-Smirnov test, we find that ET-LV galaxies are, on average, less massive than ET-HV at $>99\%$ 
confidence level. This tendency is in agreement with \citet{Gill05} who found that backsplash galaxies lose part of their mass in a 
single passage through the cluster centre. Nevertheless, it should be notted that \citet{Gill05}
used DM simulations and we are considering stellar masses.

\begin{figure}[t]
   \centering
   \includegraphics[width=9cm, angle=270]{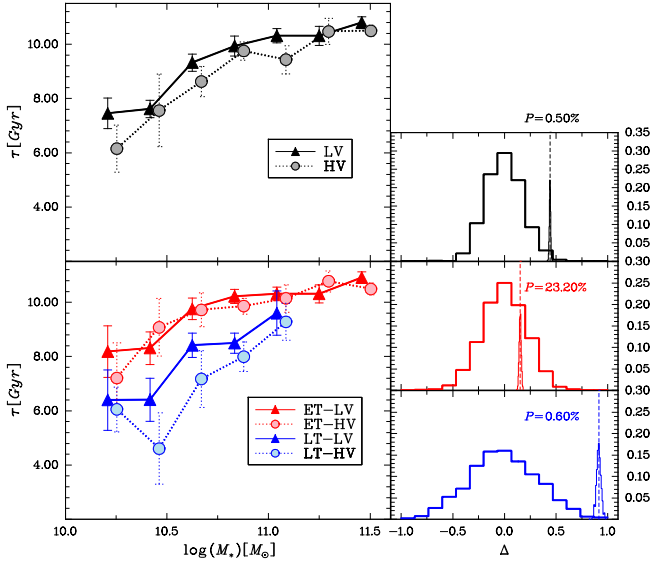}
    \caption{Age as a function of stellar mass. Points represent the median in each bin, vertical error-bars 
are obtained by using the bootstrap resampling technique. Filled triangles and 
continuous lines represent the population of LV galaxies while HV objects are shown as filled circles 
and dotted lines. The \emph{upper left} panel corresponds to the total sample of galaxies, while early- (red) and
late-type (blue) galaxies are shown in the \emph{bottom left} panel. HV galaxies have been shifted in $0.03$ on the $x$-axis. 
\emph{Right} side panels show the confidence test (see text). Vertical dash line show ${\Delta}_{obs}$ value
obtained by eq. \ref{eq:rara}.} 
\label{fig:age}
\end{figure}

Given that galaxy properties strongly depend on stellar mass \citep{Kauff:2003} and that LV and HV samples have 
different stellar mass distributions, we explore the dependence of 
age, metallicity and the SFF as a function of the stellar mass for LV and HV galaxies. 
Fig. \ref{fig:age} shows the median of age $\tau$ as a function of the stellar mass $\log(M_{\ast})$.
Error bars are obtained 
by the bootstrap re-sampling technique. The upper left panel corresponds
to the total sample while early and late are shown in the lower left panel. 
For the total sample (LT+ET) we find that LV galaxies are older than HV, however this tendency is non 
statistically significant. If morphological types are considered, we can see that age difference is 
only present for LT galaxies.
Although for most of the bins the differences in age as a function of the stellar mass between 
LT-LV and LT-HV are not statistically 
significant, the reported trend is systematic and appears in nearly all bins. 
Except for the contamination produced by the error in the mass determination, each bin
of stellar mass is independent, therefore, the systematic differences give confidence to the trend. 
In order to quantify the significance of this effect, we apply
the following test. We compute the parameter ${\Delta}_{obs}$ as follow:

\begin{equation}
\label{eq:rara} 
\Delta _{ obs }\equiv \sum _{ k=1 }^{ N_{ bin } } \sum _{ i=1 }^{ N_{ A }^{ k } } 
\sum _{ j=1 }^{ N_{ B }^{ k } } (Y_{ A_{ i } }^{ k }-Y_{ B_{ j } }^{ k })/
\sum _{ k=1 }^{ N_{ bin } } N_{ A }^{ k }N_{ B }^{ k }
\end{equation}

where $Y_{ A_{ i } }^{ k }$ is the value of the property $Y$ (e.g. the age $\tau$) and
$A$ denotes the sample (e.g. LV) of the $ith$ galaxy of a total $ N_{ A }^{ k }$  objects in the kth bin of stellar mass. 
$B$ denotes the second sample (e.g. HV). Basically,  ${ \Delta  }_{ obs }$ represents the mean difference
between two samples of galaxies once the trend of the galaxy property with the stellar mass is removed. 
The resulting values of  ${ \Delta  }_{ obs }$ for age are shown as vertical lines in the right-side 
panels of Fig. \ref{fig:age}. In order to test how significant these values are, we implement the following test.
For each bin of stellar mass, we create a sample of $ N_{C}^{ k } $ galaxies composed of the sum of the 
$ N_{A }^{ k } +  N_{B }^{ k }$ galaxies in the kth bin. From $ N_{C }^{ k }$, we randomly select 
two samples of galaxies with $ N_{A }^{ k }$ and $ N_{B }^{ k }$ galaxies and compute 
${ \Delta  }_{ boot }$ as in equation \ref{eq:rara}.  The procedure is repeated 1000 times. The resulting
values of ${ \Delta  }_{ boot }$ are shown as thick histograms in the right-side panels of Fig. \ref{fig:age}.
Finally, we compute the probability of having values of ${ \Delta  }_{ boot }$ that depart from zero more
than ${ \Delta  }_{ obs }$. These $P$ values are quoted in the corresponding panels of Fig. \ref{fig:age}. 
We can see that the differences in age between the LV and the HV samples are quite significant for the LT and the
total sample ($>$99\%). Nevertheless, for the total sample it should be noted that of the seven bins of stellar 
mass only three show differences in age, consequently the effect for the total sample should be considered as 
very marginal.  In order to understand how the uncertainties on the model-derived quantities impact 
in our results, we implement the following test.
\citet{Tojeiro10} found that the average systematic error on an 8 Gyr population is
approximately 1 Gyr. To each galaxy in our sample we compute $\tau' = \tau + \delta$ where $\delta$ is a random 
normally distributed error with 
a standard deviation of 1 Gyr. This new sample is used to compute ${ \Delta  }_{ obs }'$ using equation \ref{eq:rara} and
the procedure is repeated 1000 times. The results of this test are shown as a thin histogram in the right panels 
of Fig. \ref{fig:age}. As can be observed, the uncertainty introduced by the systematic errors in the model-derived 
quantities are an order of magnitude smaller than the uncertainties produced by the cosmic variation (galaxy-to-galaxy). 

\begin{figure}[t]
   \centering
   \includegraphics[width=4.5cm, angle=270]{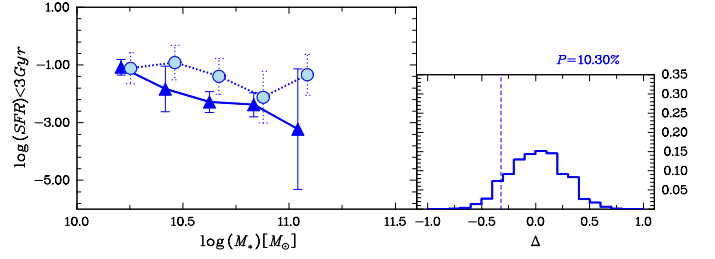}
   \caption{\emph{Left} side panels: star formation fraction during the last 3 Gyrs as a function of stellar mass for LT galaxies. 
   Points represent the median in each bin, and vertical error-bars are obtained by using the bootstrap resampling 
technique. \emph{Right} side panels show the confidence test (see text). Points and line types as in Fig. \ref{fig:age}.}
\label{fig:sff}
\end{figure}

No differences are found when the metallicity is considered.
If the fraction of star formation during the 
last 3 Gyr is considered, our analysis suggests (with a marginal significance ($\sim$90\%)) that the LT-HV galaxies 
have higher SFF than LT-LV galaxies (see Fig. \ref{fig:sff}).

\begin{figure}[t]
   \centering
   \includegraphics[width=8cm]{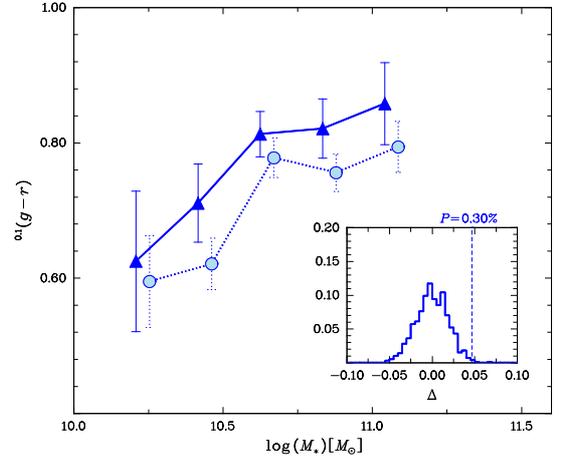}
 \caption{$^{0.1}(g-r)$ as a function of stellar mass for LT galaxies. Points 
represent the median in each bin. Vertical error-bars are obtained by using the bootstrap resampling technique. 
Inside panel show the confidence test (see text).
Symbols types as in Fig. \ref{fig:age}.}
\label{fig:gr}
\end{figure}

\begin{figure}[t]
   \centering
   \includegraphics[width=5.5cm, angle=270]{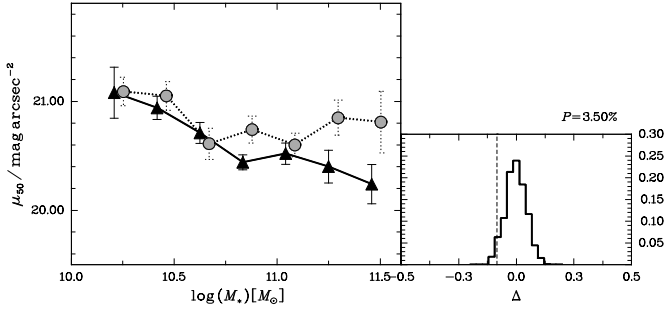}
      \caption{Surface brightness $\mu_{50}$ as a function of the stellar mass for the total sample of galaxies. 
      Points represent the median in each bin. Vertical error-bars are obtained by using the bootstrap resampling technique. 
\emph{Right} side panel show the confidence test (see text).
Symbol types as in Fig. \ref{fig:age}.}
\label{fig:mu50}
\end{figure}

\begin{figure}[t]
   \centering
    \includegraphics[width=5.5cm, angle=270]{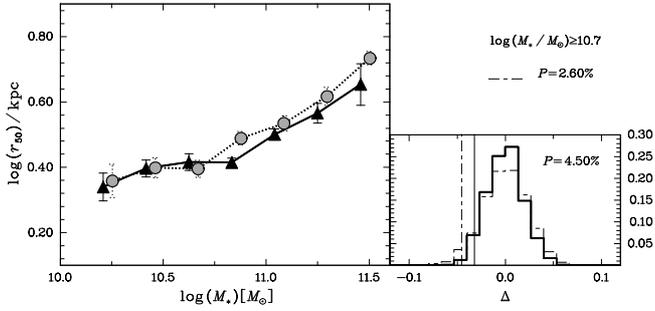}
      \caption{\emph{Left} side panels: Petrosian half-light radius $r_{50}$ as a function of stellar mass for the total sample 
      of galaxies.  Points represent 
the median in each bin. Vertical error-bars are obtained by using the bootstrap resampling technique. 
\emph{Right} side panels show the confidence test for the total sample (continuous lines) and for 
a subsample of galaxies with $\log(M_{\ast})\geqq 10.7$ (dash dotted lines). Symbol types 
as in Fig. \ref{fig:age}.}
\label{fig:r50}
\end{figure}

For the subsamples of LV and HV galaxies we also analyse the correlation between the stellar mass and several 
photometric properties. In Fig. \ref{fig:gr} we plot the median of $^{0.1}(g-r)$ as a function 
of $\log(M_{\ast})$ for late-type galaxies. We find that LT-HV galaxies are 
bluer than LT-LV ($>99\%$ confidence level). This effect is in agreement with the 
excess of recent star formation showed in 
Fig. \ref{fig:sff}. Fig. \ref{fig:mu50} shows the median of the surface brightness $\mu_{50}$ as a function of 
$\log(M_{\ast})$ for the total sample. We see 
signals at 96.5\% confidence level in the sense that LV galaxies have 
higher surface brightness than HV galaxies. 
If the mean absolute magnitude of the LV and HV galaxies as function of the stellar mass is compared, no difference 
is observed. Finally, in Figure \ref{fig:r50} 
we show the correlation between the Petrosian half-light radius $r_{50}$ as a function of $\log(M_{\ast})$ for the total 
sample of galaxies. Although marginally significant, we can see that LV galaxies with $\log(M_{\ast}/M_{\odot})\geqq 10.7$ are on 
average smaller than HV. The right panel of the figure shows
the result of our test of significance, where we also include the same analysis for a subsamples of galaxies with 
$\log(M_{\ast}/M_{\odot})\geqq 10.7$ (dash dotted lines).
The surface brightness and $r_{50}$ are not independent parameters; in fact the lack of differences in absolute magnitude
between LV and HV galaxies suggest that the differences in $\mu_{50}$ observed in Fig. \ref{fig:mu50} are due to the smaller 
mean size of LV than that of HV galaxies.

\begin{figure}[t]
   \centering
    \includegraphics[width=5.5cm]{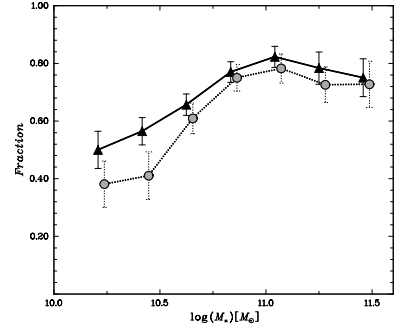}
      \caption{Fraction of ETs as a function of the stellar mass. Points represent 
the median in each bin. Vertical error-bars are obtained by using the bootstrap resampling technique. Symbol types as in Fig. \ref{fig:age}.}
\label{fig:frac}
\end{figure}

For the whole sample of LV+HV galaxies, we find
that the fraction of ETs grows with the stellar mass, going from 50\% for the lowest mass bin up to 
80\% for high mass galaxies. We test whether these fractions depend on the subsample of galaxies. We 
find that, for all the mass bins considered in this work, the fraction of ETs is larger in the LV sample 
than in the HV (see Fig. \ref{fig:frac}). The difference is larger for low mass galaxies, where 
the fraction of ETs in the LV 
sample is up to 35\% larger than in the HV sample. For the high mass bins, this difference falls to 5\%. 
Using the same test previously discussed, we find that the LV and HV samples have different fraction of 
ETs at 98\% confidence level.

\section{Conclusions and Discussion}
\label{sec:conclu}

Based on a sample of galaxy clusters composed of 90 X-ray selected and MaxBCG clusters in the redshift 
range $0.05<z<0.10$, we analyse properties of 1060 SDSS galaxies on the outskirt of clusters 
($1 < r/r_{vir} < 2$). Using the radial velocity of galaxies (relative to the mean cluster velocity), 
we select low and high velocity subsamples as follows: $|v_{los}| < 0.5\sigma$  and 
$|v_{los}| > \sigma$ respectively. According to theoretical predictions, a significant fraction of 
the first subsample  should be backsplash galaxies, that is, objects that have already orbited near 
the cluster centre. In the backsplash scenario, the sample of galaxies with high relative velocity (HV) 
would include a significant fraction of objects that are falling into the cluster for the first time. 
We analyse two sets of properties. i) Photometric 
properties: the colour $^{0.1}(g-r)$, the Petrosian $r$-band absolute magnitude $M_{^{0.1}r}$, 
the radius that encloses 50\% of the Petrosian flux and
the $r$-band surface brightness computed inside $r_{50}$. ii) Properties derived from the spectral 
synthesis analysis: we use the 
VESPA algorithm (\citealt{Tojeiro09}) to compute the stellar mass $M_{*}$, 
the mean age $\tau$ weighted by mass, the metallicity $Z$ and the fraction of the stars that were 
formed during the last 3 Gyrs.



We find that ET-LV galaxies are systematically less massive than ET-HV, in agreement with the predictions 
of \citet{Gill05} who found clear evidences that 
LV galaxies lose part of their mass in a single passage through the cluster centre. 
For the whole sample of galaxies, we find that
the fraction of ETs grows with the stellar mass, going from
50\% for the lowest mass bin and up to 80\% for high mass galaxies.
The fraction of ETs was larger in the LV sample
than in the HV. The difference was larger for low mass galaxies,
in which the fraction of ETs in the LV sample was up to
35\% larger than in the HV sample. This result supports the 
scenario in which the central region of clusters of galaxies can contribute to the transformation of 
morphological types.

Galaxies with the same stellar mass and in the same environment are expected to be similar. 
Departures from the mean properties may be directly linked to differences in the recent dynamical history, 
as would be the case when comparing LV and HV galaxies with the same stellar mass in the same cluster region. 
We test this hypothesis for each of the galaxy properties considered in our work. 
Our results show that, for the same stellar mass, 
LV galaxies are on average older than HV.  If morphological types are considered, we find that 
this effect corresponds to LT galaxies. These results are
consistent with the age gradients found in clusters by \citet{Rakos:2007}. 
\citet{Smith:2012} found that the age gradient depends on the mass of galaxy being lower for the most 
massive objects. These authors suggested that, even if there is a significant pre-processing in 
groups, the location in the cluster core at z = 0 effectively selects galaxies which passed 
through all stages of the environmental history, including pre-processing, earlier than those 
which today reside in the cluster outskirts (see also  \citealt{Price:2011}). Within this scenario, 
LV are more 
advanced in the process of accretion and should therefore be, on average older, as observed in our 
analysis. 

We find some evidences that LT-HV galaxies have formed more stars 
during the last 3 Gyrs than LT-LV galaxies, consistent with \citet{Mahajan11} and with models that assume 
that a passage through the central region 
of clusters tends to quench star formation due to processes like ram pressure stripping or 
strangulation. This phenomenon is also reflected in the colours of galaxies. Our results show that 
LT-LV galaxies are less blue (have had less recent star formation) than LT-HV.

Although marginally significant, we find that LV galaxies with $\log(M_{\ast}/M_{\odot})\geqq 10.7$ 
tend to have a higher surface brightness than HV. 
Analysing the size and the luminosity of galaxies, our results suggest that the higher values of $\mu_{50}$ 
for LV galaxies is likely to be due to smaller size of LV galaxies (for a given stellar mass) in comparison to 
HV galaxies. These results could be related to  
gravitational type processes, in which galaxies that have orbited the central regions of clusters 
are more likely to suffer tidal effects, such as the loss of part of their mass as well as a 
re-distribution of matter towards more compact configurations. This result is in agreement with
\citet{Aguilar86} who using N-body simulations found that strong collisions result in a reduced effective radius 
and a higher surface density.

\begin{acknowledgements}
This work has been partially supported with grants from Consejo Nacional
de Investigaciones Cient\'\i ficas y T\'ecnicas de la Rep\'ublica Argentina
(CONICET) and Secretar\'\i a de Ciencia y Tecnolog\'\i a de la Universidad 
de C\'ordoba. We thank the anonymous referee for a careful reading and useful 
questions and suggestions that improved the presentation of this paper. We would
like to thank Marcelo Lares for helpful discussions about the error analysis.
Funding for the Sloan Digital Sky Survey (SDSS) has been provided by the 
Alfred P. Sloan 
Foundation, the Participating Institutions, the National Aeronautics and Space 
Administration, the National Science Foundation, the U.S. Department of Energy, 
the Japanese Monbukagakusho, and the Max Planck Society. The SDSS Web site is 
http://www.sdss.org/.
The SDSS is managed by the Astrophysical Research Consortium (ARC) for the 
Participating Institutions. The Participating Institutions are The University 
of Chicago, Fermilab, the Institute for Advanced Study, the Japan Participation 
Group, The Johns Hopkins University, the Korean Scientist Group, Los Alamos 
National Laboratory, the Max Planck Institut f\"ur Astronomie (MPIA), the 
Max Planck Institut f\"ur Astrophysik (MPA), New Mexico State University, 
University of Pittsburgh, University of Portsmouth, Princeton University, 
the United States Naval Observatory, and the University of Washington.
\end{acknowledgements}


   
 \end{document}